\documentclass[nofootinbib,prd,aps,twocolumn,showpacs,tightenlines,floatfix]{revtex4}

\usepackage{graphics,bm}
\usepackage{epsfig}
\usepackage{graphicx}

\newcommand{\beq}{\begin{equation}}
\newcommand{\eeq}{\end{equation}}
\newcommand{\bqa}{\begin{eqnarray}}
\newcommand{\eqa}{\end{eqnarray}}

\begin{document}

\pagestyle{plain}
\font\tenrm=cmr10
\def\sumint{\hbox{$\sum$}\!\!\!\!\!\!\int}
\def\square{\vcenter{\vbox{\hrule height.4pt
          \hbox{\vrule width.4pt height4pt
          \kern4pt\vrule width.3pt}\hrule height.4pt}}}
\def\boxx{\square}

\title{Relativistic Bose gases at finite density}
\author{Jens O. Andersen} 
\affiliation{Nordita,\\ \\Blegdamsvej 17, DK-2100 Copenhagen {\O}, Denmark}

\date{\today}
\begin{abstract}
We consider a massive relativistic Bose gas 
with $N$ complex scalars at finite density.
At zero temperature, we calculate the pressure, charge density and the
speed of sound in the one-loop approximation.
In the nonrelativistic limit, we obtain the 
classic results for the dilute Bose gas.
We also discuss finite-temperature effects. In particular,
we consider the problem of calculating the critical temperature
for Bose-Einstein condensation.
Dimensional reduction and effective-field-theory methods are used 
to perturbatively calculate the effects of the nonstatic Matsubara
modes. Calculations of $T_c$ in the effective 
3d theory require nonperturbative methods.
Using the Monte Carlo simulations of 
X. Sun [Phys. Rev. {\bf E67}, 066702 (2003)]
and the seven-loop variational perturbation theory (VPT)
calculations of B. Kastening 
[Phys. Rev. {\bf A70}, 043621 (2004)], we obtain $T_c$
for $N=2$ to second order in the interaction.  
\end{abstract}

\maketitle

\small

\section{Introduction}
The realization of Bose-Einstein condensation
(BEC) of trapped alkali atoms 
almost ten years ago
has created an enormous interest in the properties
of the weakly interacting Bose gas~\cite{pethick,dalfovo,jens2}. 
The temperature at which these systems Bose condense is of the order
$10^{-5}$ Kelvin, which is many orders of magnitude higher than the
condensation temperature and the onset of superfluidity in $^4$He (2.17K). 
Both trapped alkali gases and $^4$He are examples of nonrelativistic
systems.

BEC in relativistic Bose systems typically takes place in matter under
extreme conditions. 
For example, kaons may condense in the color-flavor
locked phase of high-density QCD~\cite{shot,sjafer,sjafer2}.
This phase is a superconducting phase of QCD 
which arises from an instability of the Fermi surface; In analogy
with ordinary BCS-theory, a weak attraction among the quarks
in one or more channels
results in the formation of Cooper pairs and the spontaneous breakdown of
the color symmetry of QCD~\cite{color1,color2,color3}.
Such a phase may be found in the interior of compact stars if
the density is sufficiently high.
The linear $SU(2)_L\times SU(2)_R$-symmetric sigma model at
finite chemical potential $\mu$ for the hypercharge 
is used as a toy model for the description of kaon 
condensation in the color-flavor locked phase of QCD~\cite{igor1,igor2}.

Calculation of the critical temperature or the critical
density for Bose-Einstein condensation 
of an ideal Bose gas has been a standard text-book calculation
for a number of years~\cite{kap}.
The charge density $n$ 
of excited bosons as a function of temperature is given by
\bqa
n&=&\int{d^3p\over(2\pi)^3}\bigg[{1\over e^{\beta\left(\omega-\mu\right)}-1}
-{1\over e^{\beta\left(\omega+\mu\right)}-1}
\bigg]\;,
\label{ndef}
\eqa
where $\omega=\sqrt{p^2+m^2}$, $\beta=1/T$ and $\mu$ is the chemical
potential. We have set $\hbar=k_B=1$.
Bose-Einstein condensation takes place when the chemical
potential is equal to the mass $m$ of the bosons.
At that temperature, all the charge can no longer be accomodated in the
excited states and it condenses into the ground state.
Generally, $T_c$ can only be calculated numerically, but 
in various limits analytic results can be obtained.
For example, 
in the nonrelativistic (NR) limit, the critical density $n_c$
calculated from 
Eq.~(\ref{ndef}) is 
\bqa
n_c&&=\zeta\left(\mbox{${3\over2}$}\right)
\left({mT\over2\pi}\right)^{3/2}\;.
\label{nr}
\eqa
Inverting this equation, one obtains the well-known result
$T_c=2\pi/m\left[n/\zeta\left(\mbox{${3\over2}$}\right)\right]^{2/3}$ .
Similarly, for $m\rightarrow0$, one finds
\bqa
T_c&=&\left({3n\over m}\right)^{1/2}\;.
\label{tc00}
\eqa
Thus, in the ultrarelativistic limit $m=0$, $T_c$ is infinite, or equivalently,
the critical charge density is zero. All charge resides in the 
ground state irrespective of the temperature.

In the context of nonrelativistic field theory, the problem of calculating
the transition temperature for Bose-Einstein condensation 
with a weak interaction has 
a very long history and conflicting results have appeared in the 
literature~\cite{jens2}.
A first-order perturbative calculation gives no correction to the
ideal-gas result, while higher-order calculations are
plagued with infrared divergences. This is a typical example of
infrared divergences that arise in the vicinity of a second-order
phase transition. The long-distance physics in the critical region
is nonperturbative and one has to sum up an infinite set of diagrams
to obtain a finite result. 
The problem was solved only recently
by Baym {\it et al.}~\cite{gordon} who realized that it
can be reduced to nonperturbative calculations using a classical 
three-dimensional field theory.
Once it was understood how to organize the problem, the calculation of 
$T_c$ has later been carried out using several methods. These include 
$1/N$ techniques~\cite{gordon1,toma},
lattice
simulations~\cite{arnold,svis}, the linear delta 
expansion~\cite{eric,kneur}, variational perturbation 
theory~\cite{kleinert,boris,boris2}, and renormalization group 
methods~\cite{kopietz,blai}. 
In particular, the use of effective field theory methods to obtain an 
effective three-dimensional
field theory combined with high precision lattice calculations has settled 
the issue in a very elegant way~\cite{arnold,arnold2}.

The $O(2)$-symmetric
relativistic Bose gas at finite temperature and chemical potential
has been studied in detail by Benson, Bernstein, and 
Dodelson~\cite{dodel1,dodel2}, while the problem of calculating 
$T_c$ was addressed in a paper by Bedingham and Evans employing the
linear delta expansion~\cite{evans}. In the present paper, we 
examine the relativistic Bose gas in a more general setting where 
we consider $N$ coupled scalars.
For $N=1$, it reduces to the
standard case, while for $N=2$, the system is isomorphic to the
linear $SU(2)_L\times SU(2)_R$-symmetric sigma model and hence relevant for
kaon condensation in stars.

\section{Perturbation theory}
In this section, we briefly discuss the perturbative framework
for a massive Bose gas with $N$ charged scalars 
at finite chemical potential $\mu$. The action is
\bqa\nonumber
S&=&\int\;dt\int\;d^3x
\bigg[(\partial_0+i\mu)\Phi^{\dagger}(\partial_0-i\mu)\Phi
\\ &&
-\left({\partial_i}\Phi^{\dagger}\right)\left(
{\partial_i}\Phi\right)
-m^2\Phi^{\dagger}\Phi-{\lambda}\left(\Phi^{\dagger}\Phi\right)^2
\bigg]
\;,
\label{lag}
\eqa
where $\Phi=(\Phi_1,\Phi_2...,\Phi_N)$ and $\Phi_i$ is a complex
scalar field.
We first parametrize the quantum field $\Phi_1$ in terms of
a time-independent vacuum expectation value $\phi_0$ and two real
quantum fluctuating fields:
\bqa
\Phi_1&=&\phi_0+{1\over\sqrt{2}}\left(\phi_1+i\phi_2\right)\;.
\label{break}
\eqa
Similarly, the remaining complex fields $\Phi_2,...,\Phi_N$ are parametrized
in terms of $2N-2$ real fields $\phi_3,...,\phi_{2N}$.
Substituting Eq.~(\ref{break}) into Eq.~(\ref{lag}),
the action can be written as
\bqa
S&=&S_0+S_{\rm free}+S_{\rm int}\;.
\eqa
where 
\bqa
S_0&=&\int\;dt\int\;d^3x
\left[\left(\mu^2-m^2\right)\phi_0^2-\lambda\phi_0^4\right]\;,
\\ \nonumber
S_{\rm free}&=&\int\;dt\int\;d^3x
\Bigg\{{1\over2}\phi_i\bigg[-{\partial^2\over\partial t^2}
+\nabla^2+\mu^2-m^2
\\&&\nonumber
-2\lambda\phi_0^2-4\delta_{j1}\delta_{j1}\lambda\phi_0^2
\bigg]\phi_i
+i\mu\left[\phi_2{\partial\phi_1\over\partial t}
-\phi_1{\partial\phi_2\over\partial t}
\right.\\ && \left.
+...+
\phi_{2N}{\partial\phi_{2N-1}\over\partial t}
-\phi_{2N-1}{\partial\phi_{2N}\over\partial t}
\right]
\Bigg\}\;,
\\
\nonumber
S_{\rm int}&=&-\int\;dt\int\;d^3x\bigg[
\sqrt{2}\left(m^2-\mu^2+2\lambda\phi_0^2\right)\phi_1\phi_0
\\ && 
\hspace{-0.4cm}
+\sqrt{2}\lambda
\left
(\phi_1^2+\phi_2^2+...+\phi_{2N}^2\right)\phi_1\phi_0
+{1\over4}{\lambda}\left(\phi_i\phi_i\right)^2\bigg]\;.
\eqa
The propagators that correspond to the free part $S_{\rm free}$
of the action are given by
\begin{widetext}
\bqa
D_1(\omega,p)&=&{i\over(\omega^2-\omega_{1+}^{2})(\omega^2-\omega_{1-}^{2})}
\left(\begin{array}{cc}
\omega^2-p^2-m_1^2&2i\mu\omega\vspace{2mm}
\\
-2i\mu\omega&\omega^2-p^2-m_2^2
\end{array}\right)\;, \\ \nonumber
&& \\ \nonumber
&& \\
D_2(\omega,p)&=&{i\over(\omega^2-\omega_{2+}^2)(\omega^2-\omega_{2-}^2)}
\left(\begin{array}{cc}
\omega^2-p^2-m_2^2&2i\mu\omega\vspace{2mm}
\\
-2i\mu\omega&\omega^2-p^2-m_2^2
\end{array}\right)\;,
\eqa
where the dispersion relations are
\bqa
\omega_{1\pm}(p)&=&
\sqrt{p^2+2\mu^2+{1\over2}\left(m_1^2+m^2_2\right)
\pm{1\over2}\sqrt{(m_1^2+m_2^2)^2+2\mu\left(2\mu^2+m_1^2+m_2^2\right)
+4\mu^2p^2}}\;,
\label{disp0} 
\\
\omega_{2\pm}(p)&=&\sqrt{p^2+\mu^2+m_2^2}\pm\mu
\label{disp0p}
\;.
\eqa
\end{widetext}
Here the tree-level masses $m_1^2$ and $m_2^2$ are
\bqa
m_1^2&=&-\mu^2+m^2+6{\lambda}\phi_0^2\;,\\
m_2^2&=&-\mu^2+m^2+2{\lambda}\phi_0^2\;.
\eqa
In the minimum of the classical action,
$m_1^2=2(\mu^2-m^2)$ and $m^2_2=0$, and so
the dispersion relations reduce to
\bqa\nonumber
\hspace{-1.2cm}
\omega_{1\pm}(p)&=&\sqrt{p^2+3\mu^2-m^2\pm\sqrt{(3\mu^2-m^2)^2+4\mu^2p^2}}\;,
\label{disp1}
\\ &&
\\
\omega_{2\pm}(p)&=&\sqrt{p^2+\mu^2}\pm\mu
\;.
\label{disp2}
\eqa
From these equations, we see that there are two massless modes
that in the long-wavelength behave as
\bqa
\label{l1}
\omega_{1-}(p)&=&\sqrt{{\mu^2-m^2\over3\mu^2-m^2}}\;p\;,\\ 
\omega_{2-}(p)&=&
{p^2\over2\mu}\;.
\label{l2}
\eqa
The other excitations $\omega_1^+(p)$ and $\omega_2^+(p)$
are gapped with gaps $\Delta_1=\sqrt{2(3\mu^2-m^2)}$ and
$\Delta_2=2\mu$.
In the case $N=2$, 
the gapless particles $\omega_1^-$ 
and $\omega_2^-$ carry the 
quantum numbers of $K^+$ and $K^0$, while the massive modes 
$\omega_1^+$  and $\omega_2^+$  carry those of $K^-$ and 
$\bar{K}^0$~\cite{igor1,igor2}.
Note that there are only $N$ massless modes despite the fact that the potential
has $2N-1$ flat directions which also is the number of broken generators.
This is in agreement with the counting rule derived by Nielsen and
Chadba~\cite{holger}, which states that the modes with a quadratic
dispersion relation must be counted twice.
Secondly, due to the quadratic dispersion relation for small $p$,
the Landau criterion~\cite{landau} for superfluidity can never be satisfied,
except for $N=1$.
Thus despite the presence of a Bose condensate, the system is not a 
superfluid.

\section{zero temperature}
In this section, we apply perturbation theory at zero temperature
to calculate the pressure, charge density, and the speed of sound in 
the one-loop approximation.

The partition function ${\cal Z}$ is given by the path integral
\bqa
{\cal Z}&=&\int{\cal D}\Phi^{\dagger}{\cal D}\Phi\;
e^{iS}
\eqa
where the action $S$ is given by~(\ref{lag}).
The pressure ${\cal P}$ is 
\bqa
{\cal P}(\mu)&=&-i\;{\ln{\cal Z}\over VT}\;,
\eqa
where $VT$ is the space-time volume of the system.
The charge density can be found by differenting the pressure with 
respect to $\mu$:
\bqa
n(\mu)&=&{\partial{\cal P}(\mu)\over\partial\mu}\;.
\label{density}
\eqa
The speed of sound $c$ is given by the coefficient of $\omega_{1-}(p)$
as $p\rightarrow0$.
Corrections to the tree-level result can be found by 
calculating the dispersion relation in the long-wavelength limit
including the self-energy 
function $\Pi_{1-}(\omega,p)$.
It can also be derived once we know the charge density and is given by
\bqa
c^2={n\over\mu}{\partial\mu\over\partial n}\;.
\label{deri}
\eqa
The chemical potential measures the amount of energy needed to add a particle
to the system, and in the nonrelativistic limit, we introduce the
nonrelativstic chemical potential $\mu_{\rm NR}$ by $\mu=m+\mu_{\rm NR}$.
In the NR limit, Eq.~(\ref{deri}) is therefore replaced by
\bqa
c^2={n\over m}{\partial \mu_{\rm NR}\over\partial n}
\eqa

\subsection{Pressure}
The mean-field pressure ${\cal P}_0$ is found by evaluating minus
the classical 
thermodynamic potential $\Omega_0(\mu,\phi_0)$ at the minimum of 
the classical action $S_0$:
\bqa
{\cal P}_0(\mu)&=&{1\over4\lambda}\left(\mu^2-m^2\right)^2\;.
\eqa
The one-loop contribution to the effective potential is
\bqa\nonumber
\Omega_1(\mu,\phi_0)
&=&{1\over2}i\int{d\omega\over2\pi}\int_{\bf p}
\big[\ln\det D_1(\omega,p)
\\ \nonumber&&
+(N-1)\ln\det D_2(\omega,p)\big]
+\Delta_1m^2\phi_0^2
\\ &&
+\Delta_1\lambda\phi_0^4
+\Delta_1{\cal E}\;,
\eqa
where $\Delta_1m^2$, $\Delta_1\lambda$, and
$\Delta_1{\cal E}$ are the one-loop 
mass counterterm, coupling constant counterterm, and
vacuum counterterm, respectively.
After integrating over the energy $\omega$, we obtain
\bqa\nonumber
\Omega_1(\mu,\phi_0)&=&
{1\over2}
\int_{\bf p}\big[
\omega_{1\pm}(p)+(N-1)\omega_{2\pm}(p)\big]
+\Delta_1m^2\phi_0^2
\\ &&
+\Delta_1\lambda\phi_0^4
+\Delta_1{\cal E}\;.
\eqa
The integral involving $\omega_{2\pm}$ can be calculated
analytically in dimensional regularization, but the integral of
$\omega_{1\pm}$ cannot. In order to extract the divergences analytically,
we make subtractions in the integrand that render the integral finite
in $d=3$ dimensions and then extract the poles in $d-3$ from the subtracted
integrals. 
The substraction term $\Omega_{\rm sub}$
should not introduce any infrared divergences. Our choice for the
subtracted integral is
\bqa\nonumber
\Omega_{\rm sub}&=&
\int_{\bf p}
\left[p+{m^2+4\lambda\phi_0^2\over2p}
-{m^4+8m^2\lambda\phi_0^2+20\lambda^2\phi_0^4
\over8(p^2+\mu^2)^{3/2}}\right]\;.
\\ &&
\eqa
The first two terms in ${\Omega}_{\rm sub}$ vanish identically in 
dimensional regularization since there is no mass scale in the
integrand. The last term is given in Eq.~(\ref{last}).
The one-loop thermodynamics potential can then be written as 
\bqa\nonumber
\Omega_1&=&-{1\over2(4\pi)^2}\bigg\{m^4\left[
N\left({1\over\epsilon}+2L\right)+{3\over2}(N-1)
\right]
\\ \nonumber&&
+4{m^2\lambda\phi_0^2}\left[\left(N+1\right)
\left({1\over\epsilon}+2L\right)
+{3\over2}(N-1)
\right]
\\ && \nonumber
+4\lambda^2\phi_0^4\left[
\left(N+4\right)\left({1\over\epsilon}+2L\right)
+{3\over2}(N-1)\right]
\bigg\}
\\ &&
+\Delta_1m^2\phi_0^2
+{\Delta_1\lambda}\phi_0^4
+\Delta_1{\cal E}
\;,
\eqa
where $L=\ln\left({\Lambda\over\mu}\right)$ and
$g$  is a function of the ratio $m/\mu$ that must be evaluated
numerically:
\bqa
g(m/\mu)
&=&{1\over2}\int_{\bf p}
\omega_{1\pm}-\Omega_{\rm sub}
\;.
\eqa
The counterterms necessary to cancel the poles in $\epsilon$ 
are~\cite{kleinert}:
\bqa
\Delta_1{\cal E}&=&{Nm^4\over2(4\pi)^2\epsilon}\;,\\
\Delta_1m^2&=&{2(N+1)m^2\lambda\over(4\pi)^2\epsilon}\;,\\
\Delta_1\lambda&=&{2(N+4)\lambda^2\over(4\pi)^2\epsilon}\;.
\eqa
The one-loop contribution to the pressure ${\cal P}_1$ 
is given by $-\Omega_1$ evaluated at the classical minimum. 
After renormalization, the pressure through one loop reduces to
\bqa\nonumber
{\cal P}_{0+1}(\mu)&=&
{1\over4\lambda}\left(\mu^2-m^2\right)^2
+{1\over(4\pi)^2}\Bigg\{2m^4L
-6m^2\mu^2L
\\ &&
\hspace{-1cm}
+\mu^4\left[(N+4)L+{3\over4}(N-1)\right]
\Bigg\}
+\mu^4g(m/\mu)
\;.
\label{free}
\eqa

We next consider the NR limit of the pressure.
In this limit, $\mu_{\rm NR}\ll m$. Moreover, the kinetic energy is much
smaller than $m$ and so we can expand physical quantities in powers
of the dimensionless quantities $\mu_{\rm NR}/m$ and $k^2/2m^2$.
This yields
\bqa
\omega_{1-}^2(p)&=&{p^2\over4m^2}\left(p^2+4m\mu_{\rm NR}\right)\;,
\\\omega_{2-}^2(p)&=&{p^4\over4m^2}\;.
\label{nrdisp}
\eqa
The other quasiparticle excitations have $\omega_{1+}=\omega_{2+}=2m$ 
and so their
contribution can be neglected. In NR field theory, it is customary to
set $2m=1$ and we will do so in the remainder of this section.
Introducing the scattering length $a=\lambda/8\pi m=\lambda/4\pi$,
the pressure becomes
\bqa
{\cal P}_{0+1}&=&{\mu_{\rm NR}^2\over16\pi a}-{1\over2}
\int_{\bf p}\big[\omega_{1-}+(N-1)\omega_{2-}\big]
\;.
\eqa
Note that $\omega_{2-}$ does not contribute to the pressure since
there is there is no scale in the integral and so it set to zero
in dimensional regularization. Using a simple ultraviolet cutoff 
$\Lambda$ to
regulate the integral, the divergence would be cancelled by
a vacuum counterterm~\footnote{In the NR limit, all the divergences
at the one-loop level are power divergences and hence the parameters
require no renormalization if one uses dimensional regularization.
See Ref.~\cite{jens2} for a thorough discussion.}.
Using Eq.~(\ref{f0}), we obtain
\bqa
{\cal P}_{0+1}&=&
{\mu^2_{\rm NR}\over16\pi a}\left[1-{32\sqrt{2\mu_{\rm NR} a^2}\over15\pi}
\right]\;.
\label{pnr}
\eqa

\subsection{Charge density and the speed of sound}
We next consider the speed of sound, which is given by Eq.~(\ref{deri}).
For simplicity we consider only the ultrarelativistic and nonrelativistic
limits. In these limits, the tree-level results are $c=1/\sqrt{3}$
and $c=\sqrt{2\mu}$, respectively.

The charge density can be calculated
using Eqs.~(\ref{density}) and~(\ref{free}):
\bqa\nonumber
n={\mu^3\over\lambda}\left\{1+{\lambda\over(4\pi^2)^2}
\left[4(N+4)L+2N-7+64\pi^2g(0)\right]\right\}\;.
\\ &&
\label{den}
\eqa
Inverting~(\ref{den}) and using (\ref{deri}), we obtain 
the speed of sound due to interactions in the medium:
\bqa
c&=&{1\over\sqrt{3}}\left[1+{(N+4)\lambda\over24\pi^2}\right]\;.
\eqa
The sign of the correction is determined by the beta-function.

In the NR limit, the charge density follows from Eqs.~(\ref{density})
and~(\ref{pnr}):
\bqa
n&=&{\mu_{\rm NR}\over8\pi a}\left[
1-{8\sqrt{2\mu_{\rm NR} a^2}\over3\pi}
\right]\;.
\eqa
The speed of sound then becomes
\bqa
c&=&4\sqrt{\pi a n}\left[
1+8\sqrt{na^3\over\pi}
\right]\;,
\label{nrspeed}
\eqa
where we have eliminated $\mu_{|rm NR}$ in favor of the density $n$.
Note that the expansion parameter in the NR limit is the dimensionless
quantity
$\sqrt{na^3}$ which is referred to as the the {\it gas parameter}.
This result~(\ref{nrspeed}) 
was first derived by Beliaev~\cite{beli}, who calculated
the leading corrections to the dispersion relation~(\ref{nrdisp})
in the low-momentum limit~\footnote{In the orginal derivation, Beliaev
expressed his result in terms of the condensate density $n_0$, which
is different from the total density due to the depletion of the condensate
caused by quantum fluctuations. 
We have $n=n_0\left[1+{8\over3}\sqrt{n_0a^3\over\pi}\right]$.}

\section{Finite temperature}
We next discuss the behavior of the system 
defined by Eq.~(\ref{lag}) at finite temperature.
\subsection{Low-temperature effects}
We first consider the thermal corrections to the pressure at temperatures $T$
much lower than the chemical potential $\mu$. 
In this regime, the thermodynamics is dominated by the massless modes.
We can then approximate the dispersion relations by 
$\omega_1(p)$
and $\omega_2(p)$ by their low-momentum limits~(\ref{l1}) and~(\ref{l2}).
The pressure in the one-loop approximation is
\bqa\nonumber
{\cal P}_{0+1}&=&-{1\over2}
\sumint_P\ln\left[P_0^2+\omega_{1\pm}^2(p)\right]
\\ &&
-{1\over2}(N-1)\sumint_P\ln\left[P_0^2+\omega_{2\pm}^2(p)\right]\;.
\eqa
Omitting the contribution from the massive modes, neglecting the 
zero-temperature pieces, and using~(\ref{ft1})--(\ref{ft2}),
we obtain
\bqa
{\cal P}_{0+1}^{T}&=&{\sqrt{3}\pi^2T^4\over30}
+(N-1)T\left(\mbox{${\mu T\over2\pi}$}\right)^{3/2}
\zeta\left(\mbox{${5\over2}$}\right)
\;.
\eqa
Similarly, in the nonrelativistic limit, one finds
\bqa
{\cal P}^{T}_{0+1}&=&{\pi^2T^4\over90(2\mu_{\rm NR})^{3/2}}
+
(N-1)T\left(\mbox{${T\over4\pi}$}\right)^{3/2}
\zeta\left(\mbox{${5\over2}$}\right)\;,
\eqa
where we again have set $2m=1$. For $N=1$, this reduces to the old result
of Lee and Yang~\cite{leeyang2}.
\subsection{Dimensional reduction}
Effective field theory methods
can conveniently be used to organize the calcuation of
physical quantities whenever two or more momentum scales are well
separated. 
The condensation temperature for BEC is an ideal problem for 
applying effective field theory.
At finite temperature there are two characteristic scales in the system.
The first is the correlation length which is associated with the 
effective chemeical potential. Since the phase transition is second order,
the correlation length becomes infinite at $T_c$.
The second scale is associated with the nonzero Matsubara modes and is
of order $T$.
For distances much larger than the inverse temperature
and for temperatures sufficiently close to 
the critical temperature, 
so that the effective chemical potential is much smaller than the temperature,
the nonstatic Matsubara modes
decouple 
and the long-distance physics can be described in terms of an
effective
three-dimensional field theory for the zeroth Matsubara mode. 
The action for this effective theory is 
\bqa\nonumber
S_{\rm 3d}
&=&-\int d^3x\;\Big[
({\partial_i}\Phi^{\dagger})\left({\partial_i}\Phi\right)
+m^2_3\Phi^{\dagger}\Phi
\\ &&
+{\lambda_3}
\left(\Phi^{\dagger}\Phi\right)^2
+f+...\Big]
\;,
\label{3d}
\eqa
where the dots indicate higher-order operators.
$f$ is a referred to as the coefficient of the unit operator and
represents the contribution to the free energy density from the 
nonzero Matsubara modes.
The parameters in Eq.~(\ref{3d}) 
are functions of $T$ and the coefficients of the underlying theory~(\ref{lag}),
and are renormalized due to their coupling to the nonstatic Matsubara
frequencies. These parameters can be determined by integrating out
nonstatic modes explicitly as done by Bedingham and Evans in
Ref.~\cite{evans}, but perhaps a
more streamlined way of calculating them is by matching static Green's
functions~\cite{renorm}. 
For the purpose of matching, the chemical potential can 
formally be treated as a perturbation on the same footing as the quartic
coupling. In the effective theory $m_3^2$ is also treated as a perturbation.
The matching is carried out by calculating Green's functions 
perturbatively in
the two theories and demand they be the same for distances $R$ much larger
than $1/T$. This way of calculating static correlators introduces
infrared divergences at an intermediate stage
and must be regularized.
Note thate these
infrared divergences that appear are the same in the two theories
and hence they cancel in the matching procedure. 
We use dimensional regularization as discussed in the appendix.

\begin{figure}[htb]
\scalebox{0.6}{\includegraphics{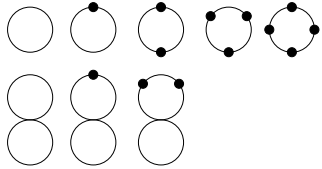}}
\caption{\label{dia}
One -and two-loop 
vacuum diagrams for $f$. A dot indicates an insertion of the
operator 
$-\left(\mu^2-m^2\right)\Phi^{\dagger}\Phi+i\mu(\Phi^{\dagger}\partial_0\Phi
-\partial_0\Phi^{\dagger}\Phi)$.}
\label{vac}
\end{figure}
In Fig.~\ref{vac}, we show the vacuum diagrams through two loops 
in the full theory and the expression is
\bqa\nonumber
{\cal F}&\approx&
N\sumint_P\ln P^2
-N\left(\mu^2-m^2\right)\sumint_P\left[{1\over P^2}-2{P_0^2\over P^4}\right]
\\ && \nonumber
-{1\over2}
N\mu^4\sumint_P\left[{1\over P^4}-8{P_0^2\over P^6}+8{P_0^4\over P^8}\right]
\\ && \nonumber
+N\mu^2m^2\sumint_P
\left[{1\over P^4}-4{P_0^2\over P^6}\right]
+N(N+1)\lambda\sumint_{PQ}{1\over P^2Q^2}
\\ && 
+2N(N+1)\lambda\mu^2\sumint_{PQ}\left[{1\over P^4Q^2}
-4{P_0^2\over P^6Q^2}\right]\;,
\label{ff}
\eqa
where we have omitted terms of order $m^4$, $\mu^6$ etc.
The sign $\approx$ is reminder that we are neglecting infrared
physics which will be taken care of by the effective theory.
Loop correction to the free energy in the 3d theory vanish since there
is no momentum scale in the loop 
integrals~\footnote{Recall that $m_3^2$ 
for the purpose of matching is treated as a perturbation
and thus the propagators are massless.}. Thus $f$ is directly
given by~(\ref{ff}).
\bqa\nonumber
f&=&-{N\pi^2\over45}T^4
\left(1-{5(N+1)\lambda\over16}\right)
\\ \nonumber&&
-{1\over6}N\mu^2T^2\left(1-{\mu^2\over4\pi^2T^2}+{(N+1)
\lambda\over8\pi^2}\right)
\\ &&
+{1\over6}Nm^2T^2\left(1-{3\mu^2\over4\pi^2T^2}\right)
\label{frr}
\eqa

\begin{figure}[htb]
\scalebox{0.5}{\includegraphics{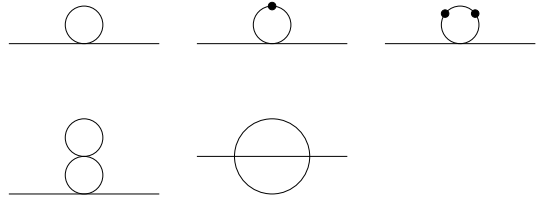}}
\caption{\label{dia2}
One -and two-loop 
Feynman diagrams for the self-energy $\Pi(p_0,{\bf p})$.
A dot indicates an insertion of the
operator $-\left(\mu^2-m^2\right)\Phi^{\dagger}\Phi+i\mu(\Phi^{\dagger}\partial_0\Phi
-\partial_0\Phi^{\dagger}\Phi)$.}
\end{figure}
The mass parameter is found by 
matching the two-point functions in the two theories
at zero external momenta.
The diagrams that contribute to the self-energy function
in the full theory through
two loops are shown in Fig.~\ref{dia2}.
The self-energy in the effective theory vanishes for the same reason
as did the loop corrections to the free energy.
This implies that the mass parameter $m_3^2$ is given directly
by evaluating the Feynman diagrams in Fig.~\ref{dia}.
We obtain
\bqa\nonumber
m_3^2&\approx&-\mu^2+m^2+2(N+1)Z_{\lambda}\lambda\sumint_P{1\over P^2} 
\\ \nonumber&&
+2(N+1)\left(\mu^2-m^2\right)\lambda\sumint\left[{1\over P^4}
-4{P_0^2\over P^6}\right]
\\ \nonumber&&
-4(N+1)^2\lambda^2\sumint_{PQ}{1\over P^2Q^4}
\\&&
-4(N+1){\lambda^2}\sumint_{PQ}{1\over P^2 Q^2(P+Q)^2}\;,
\eqa
where we again have neglected terms of higher-order terms.
$Z_{\lambda}$ is the renormalization constant for the 
coupling $\lambda$:
\bqa
Z_{\lambda}&=&1+{(N+4)\lambda\over8\pi^2\epsilon}\;.
\eqa
After renormalization, the mass term reduces to
\bqa\nonumber
m_3^2&=&-\mu^2+m^2+{(N+1)\lambda\over6}T^2\bigg[1
-{3\left(\mu^2-m^2\right)\over2\pi^2T^2}
\\ \nonumber&&
+{3\lambda\over8\pi^2}\left({1\over\epsilon}
+2-{2(N+1)\over3}\gamma
+2{\zeta^{\prime}(-1)\over\zeta(-1)}
\right.\\ &&\left.
+{(4-2N)\over3}\ln{\Lambda\over4\pi T}
\right)\bigg]\;.
\label{mrr}
\eqa
Note that the mass parameter has a UV divergence after renormalization.
The remaining divergence is exactly the one arising at the two-loop
level in the 3d effective theory (see also Sec.~\ref{tcccc}). 

Finally, we need $\lambda_3$ at the tree level. By comparing the
coefficients of the
operator $\left(\Phi^{\dagger}\Phi\right)^2$ in the two theories and taking the
different normalization of the fields into account, one finds
\bqa
\lambda_3&=&\lambda T\;.
\eqa

\subsection{Critical temperature}
\label{tcccc}
After having determined the parameters in the effective theory, the
strategy for calculating $T_c$ is as follows. First we determine the
critical chemical potential $\mu_c$ as a function of temperature and the
critical value of the $m_3^2$. Then we calculate the 
density as a function of $T$ and $\mu$ and obtain the critical 
density as a function of $T$ by the substituing the expression for
$\mu_c$. Finally, the critical temperature is determined by inverting the
critical density as a function of $T$.

The relation between the bare mass and the renormalized mass is
\bqa
m_3^2&=&m_{3,\rm ren}^2+{(N+1)\lambda_3^2\over(4\pi)^2\epsilon}\;.
\label{mbare}
\eqa
This relation is exact due to the fact that the effective 3d theory~(\ref{3d})
is superrenormalizable. 
Since the bare mass is independent of the renormalization scale,
the renormalized mass satisfies an evolution equation. This equation
relates the value of $m_{3,\rm ren}^2$
evaluated at two different normalization points $\Lambda$ and $\Lambda_0$:
\bqa
m_{3,\rm ren}^2(\Lambda_0)&=&m_{3,\rm ren}^2(\Lambda)
+{(N+1)\lambda_3^2\over4\pi^2}\ln{\Lambda_0\over\Lambda}\;.
\eqa
Using Eq.~(\ref{mbare}), Eq.~(\ref{mrr}) for the chemical potential 
becomes
\bqa\nonumber
\left(\mu^2-m^2\right)\left(1+{(N+1)\lambda\over4\pi^2}\right)&=&
{(N+1)\lambda\over6}
T^2\bigg[1
\\ && \nonumber
\hspace{-2.cm}
+{3\lambda\over8\pi^2}\left(
2-{2(N+1)\over3}\gamma+2{\zeta^{\prime}(-1)\over\zeta(-1)}
\right)
\\ && 
\hspace{-3.2cm}
+{(4-2N)\over3}\ln{\Lambda\over4\pi T}
-{16\pi^2\over(N+1)}{m_{3,\rm ren}^2(\Lambda)\over\lambda_3^2}
\bigg]\;.
\eqa
The charge density is given by
\bqa\nonumber
n&=&-\bigg\langle{\partial S_{\rm 3d}\over\partial\mu}\bigg\rangle\\
&=&-{\partial f\over\partial\mu}+
\langle\Phi^{\dagger}\Phi\rangle{\partial m_3^2\over\partial\mu}+
\lambda_3\big\langle\left(\Phi^{\dagger}\Phi\right)^2\big\rangle
\;.
\label{chd}
\eqa
The quantity $\big\langle\left(\Phi^{\dagger}\Phi\right)^2\big\rangle$
is by dimensional analysis proportional to $\lambda_3^2$.
Its contribution to the density is therefore third order in the interaction
and can be omitted in a second-order calculation.
Using Eqs.~(\ref{frr}),~(\ref{mrr}), and~(\ref{chd}), the charge density
becomes
\bqa\nonumber
n&=&{1\over3}N\mu T^2
\left[1 +{3m^2\over4\pi^2T^2}
-{\mu^2\over2\pi^2T^2}+{(N+1)\lambda\over8\pi^2
}
\right]
\\ &&
-2\mu 
\langle\Phi^{\dagger}\Phi\rangle
\;.
\label{nn}
\eqa

At the critical point, the renormalized mass
is by dimensional analysis proportional to $\lambda_3^2$.
The case $N=2$ is relevant to kaon condensates in stars and we
therefore consider this case in the following.
In the remainder of this section, we also restrict ourselves to 
the ultrarelativistic limit.
Its value was determined by Sun~\cite{sun} using lattice simulations: 
\bqa
{m_3^2(\Lambda=\lambda_3/3)\over\lambda_3^2}&=&0.002558(16)\;,
\label{mc00}
\eqa
where the renormalization scale was chosen to be $\Lambda=\lambda_3/3$.
The critical chemical potential then reduces to
\bqa\nonumber
\mu_c&=&\sqrt{{1\over2}\lambda}T\bigg[
1+{3\lambda\over(4\pi)^2}
\bigg(
-2\gamma+2{\zeta^{\prime}(-1)\over\zeta(-1)}
-0.1346\bigg)
\bigg]\;.
\\ &&
\label{myc}
\eqa

The expectation value $\langle\Phi^{\dagger}\Phi\rangle$
cannot be calculated in perturbation theory due to infrared divergences.
They depend on nonperturbative physics and can e.g. be determined 
using lattice simulations or the $1/N$-expansion.
At the critical point and for $N=2$, it was computed by Sun~\cite{sun} using
Monte Carlo calculations:
\bqa
\bigg\langle{\Phi^{\dagger}\Phi\over\lambda_3}\bigg\rangle&=&-0.00289(18)\;.
\label{mc}
\eqa
Inserting Eqs.~(\ref{myc}) and~(\ref{mc}) into Eq.~(\ref{nn}), the 
critical density becomes
\bqa\nonumber
n_c&=&\sqrt{{2\over9}\lambda}T^3\bigg[
1+{3\lambda\over(4\pi)^2}
\bigg(2
-2\gamma+2{\zeta^{\prime}(-1)\over\zeta(-1)}
+0.3217
\bigg)
\bigg]\;.
\\ &&
\label{ncf}
\eqa
Inverting this equation, we obtain the critical temperature as
a function of the density
\bqa\nonumber
T_c&=&\left({9\over2\lambda}\right)^{1/6}n^{1/3}
\bigg[1-{\lambda\over(4\pi)^2}
\bigg(2
-2\gamma+2{\zeta^{\prime}(-1)\over\zeta(-1)}
\\ &&
+0.3217
\bigg)
\bigg]\;.
\label{tcf}
\eqa
The leading-order result is the usual perturbatively calculable
high-temperature result, while the second-order 
term involves nonperturbative physics.
In contrast, the first order correction to $T_c$ in the nonrelativistic Bose
gas cannot be determined in perturbation theory~\cite{gordon}.
Note also that in accordance with Eq.~(\ref{tc00}), 
$T_c$ becomes infinite in the absence of interactions.
Finally, the impressive seven-loop VPT calculations of 
Kastening~\cite{boris,boris2}
give $0.002586(17)$  and $-0.002796(192)$
for the
quantities in Eqs.~(\ref{mc00}) 
and~(\ref{mc}).
Thus the critical temperature is within errors in complete agreement
with the lattice prediction.

\section{Summary}
In this paper, we have discussed the thermodynamics of relativistic Bose
gases at zero and finite temperature. At zero temperature, thermodynamic
quantities can be expanded in a loop expansion, and we calculated the
pressure, charge density and speed of sound in the one-loop approximation.
In the nonrelativistic limit, one easily obtains the standard results
for the dilute Bose gas.
In the critical region, perturbation theory breaks down due to infrared
divergences. However, 
one can take advantage of the fact that
there is a separation of scales to simplify the problem of calculating 
static quantities such as $T_c$.
The effects of the nonstatic modes can be caclulated perturbatively
employing dimensional reduction, while the effective three-dimensional 
theory must be treated nonperturbatively.
It is interesting to note that the leading correction to $T_c$
in the ultrarelativistic limit, is calculable in perturbation theory
while in the NR limit it is not. The reason is simply that the
chemical potential couples quadratically to $\Phi^{\dagger}\Phi$
in the first case and linearly in the latter.

By gauging the linear $SU(2)_L\times SU(2)_R$-symmetric sigma model 
in various ways, one obtains a number of interesting gauge 
theories~\cite{igor1,igor2}.  
The Higgs mechanism and the Goldstone mechanism are both realized
in the conventional manner. The interest in these models is partly due
to the fact that the rotational symmetry is broken as well and leads
to a directional dependence of the dispersion relation which 
is linear for small wave vectors and roton-like for larger wave vectors.
These models may therefore be of interest for condensed matter systems
such as superfluid helium.

\section*{Acknowledgments} 
The author would like to thank P. Arnold for useful discussions.

\appendix
\renewcommand{\theequation}{\thesection.\arabic{equation}}
\section{Formulas}
Dimensional regularization can be used to 
regularize both the ultraviolet divergences and infrared divergences
in three-dimensional integrals over momenta. 
The spatial dimension is generalized to  $d = 3-2\epsilon$ dimensions.
Integrals are evaluated at a value of $d$ for which they converge and then
analytically continued to $d=3$.
We use the integration measure
\begin{equation}
  \int_{\bf p} \;\equiv\;
  \left(\frac{e^\gamma\Lambda^2}{4\pi}\right)^\epsilon\,
  \int {d^{3-2\epsilon}p \over (2 \pi)^{3-2\epsilon}}\,.
\end{equation}
where $\Lambda$ is an arbitrary momentum scale. 
The factor $(e^\gamma/4\pi)^\epsilon$
is introduced so that, after minimal subtraction 
of the poles in $\epsilon$
due to ultraviolet divergences, $\Lambda$ coincides 
with the renormalization
scale of the $\overline{\rm MS}$ renormalization scheme.

We need the regularized integrals
\bqa
\int_{\bf p}\sqrt{p^2+\mu^2}&=&-{\mu^4\over2(4\pi)^2}\left(
{\Lambda\over\mu}\right)^{2\epsilon}
\left[{1\over\epsilon}+{3\over2}
\right]\;,
\\ 
\int_{\bf p}{1\over(p^2+\mu^2)^{3/2}}&=&{4\over(4\pi)^2}\left(
{\Lambda\over\mu}\right)^{2\epsilon}
\left[{1\over\epsilon}
\right]\;,
\label{last}
\\ 
\int_{\bf p}p\sqrt{p^2+\mu}&=&{\mu^{5/2}\over15\pi^2}
\;.
\label{f0}
\eqa

In the imaginary-time formalism for thermal field theory, 
the 4-momentum $P=(\omega_n,{\bf p})$ is Euclidean with 
$P^2=\omega_n^2+{\bf p}^2$. 
The Euclidean energy $p_0$ has discrete values:
$\omega_n=2n\pi T$ for bosons,
where $n$ is an integer. 
Loop diagrams involve sums over $\omega_n$ and integrals over ${\bf p}$. 
We define the dimensionally regularized sum-integral by
\bqa
  \hbox{$\sum$}\!\!\!\!\!\!\int_{P}& \;\equiv\; &
  \left(\frac{e^\gamma\Lambda^2}{4\pi}\right)^\epsilon\,
  T\sum_{\omega_n=2n\pi T}\:\int {d^dp \over (2 \pi)^{d}}\;.
\label{sumint-def}
\eqa
The
specific sum-integrals needed are
\bqa
\sumint_P\ln P^2
&=&-{\pi^2T^4\over45},\\ 
\sumint_P{1\over P^2}
&=&{T^2\over12}\bigg[
1+\left(
2+2{\zeta^{\prime}(-1)\over\zeta(-1)}
\right)\epsilon
\bigg]\;, \\
\sumint_P{P_0^2\over P^4}
&=&-{T^2\over24}\bigg[
1+
2{\zeta^{\prime}(-1)\over\zeta(-1)}
\epsilon
\bigg]\;, \\
\sumint_P{1\over P^4}&=&
{1 \over (4\pi)^2} \left({\mu\over4\pi T}\right)^{2\epsilon} 
\left[ {1 \over \epsilon} + 2 \gamma\right]\;,\\
\sumint_P{P_0^2\over P^6}&=&
{1 \over (4\pi)^2} \left({\mu\over4\pi T}\right)^{2\epsilon} 
{1\over4}\left[ {1 \over \epsilon} + 2+2 \gamma\right]\;,\\
\sumint_P{P_0^4\over P^8}&=&
{1 \over (4\pi)^2} \left({\mu\over4\pi T}\right)^{2\epsilon} 
{1\over8}\left[ {1 \over \epsilon} + {8\over3}+2 \gamma\right]\;,\\ 
&&\hspace{-1cm}
\sumint_{PQ}{1\over P^2Q^2(P+Q)^2}=0\;.
\label{2ll} 
\eqa
We also need to expand some sum-integrals about zero temperature.
The phonon part of the spectrum then dominates the
temperature-dependent part of the sum-integral. We can therefore approximate 
the dispersion relations
 $\omega_{1-}(p)$ and $\omega_{2-}(p)$ by 
their low-momentum limits~(\ref{l1})-(\ref{l2}),
and this gives the leading temperature correction. These are
\bqa
\sumint_P\ln\Big[P_0^2+\omega_{1-}^2(p)\Big]
&&\
\\ \nonumber&&
\hspace{-2.6cm}
=\int_{\bf p}\omega_{1-}(p)
+{T\over\pi^2}\int_0^{\infty}dp\;p^2
\ln\left[1-e^{-\beta\omega_{1-}(p)}\right]
\\
&&
\hspace{-2.6cm}=\int_{\bf p}\omega_{1-}(p)-{\pi^2T^4\over45}
\left({3\mu^2-m^2\over\mu^2-m^2}\right)^{3/2}+...
\label{ft1}
\;, \\
\sumint_P\ln[P_0^2+\omega_{2-}^2(p)\Big]&&
\\
\nonumber&&
\hspace{-2.6cm}
=\int_{\bf p}\omega_{2-}(p)
+{T\over\pi^2}\int_0^{\infty}dp\;p^2
\ln\left[1-e^{-\beta\omega_{2-}(p)}\right]
\\
&&
\hspace{-2.6cm}=\int_{\bf p}\omega_{2-}(p)-
2T\left(\mbox{${\mu T\over2\pi}$}\right)^{3/2}
\zeta\left(\mbox{${5\over2}$}\right)
+...
\label{ft2}\;.
\eqa

\end{document}